\documentclass[]{mn2e}
\usepackage{epsfig}

\title[Progenitors of SNe Ia]{The single degenerate channel for the
progenitors of type Ia supernovae}

\author[Han \mbox{\rm\&} Podsiadlowski]
{Z.~Han$^1$\thanks{E-mail: zhanwen@public.km.yn.cn}, Ph.~Podsiadlowski$^2$\\
\it
$^1$ National Astronomical Observatories / Yunnan Observatory, 
the Chinese Academy of Sciences, P.O.Box 110, Kunming, 650011, China\\
$^2$ University of Oxford, Department of Astrophysics, Keble Road, Oxford,
OX1 3RH
}
\newcommand{\Ms}{\mbox{\,M$_{\odot}$}}
\begin{document}
\maketitle

\begin{abstract}
We have carried out a detailed study of one of the most popular
evolutionary channels for the production of Type Ia supernova (SN Ia)
progenitors, the semi-degenerate channel (CO+MS), where a
carbon/oxygen (CO) white dwarf (WD) accretes matter from an unevolved
or slightly evolved non-degenerate star until it reaches the
Chandrasekhar mass limit.  Employing Eggleton's stellar evolution code
and adopting the prescription of Hachisu et al.\ (1999) for the
accretion efficiency, we have carried out full binary evolution
calculations for about 2300 close WD binary systems and mapped out the
initial parameters in the orbital period -- secondary mass
($P$--$M_2$) plane (for a range of WD masses) which lead to a
successful Type Ia supernova.  We obtained accurate, analytical
fitting formulae to describe this parameter range which can be used for
binary population synthesis (BPS) studies. The contours in the
$P$--$M_2$ plane differ from those obtained by Hachisu et al.\
(1999) for low-mass CO WDs, which are more common than massive CO
WDs. We show that white dwarfs with a mass as low as $0.67M_\odot$ can
accrete efficiently and reach the Chandrasekhar limit.  We have
implemented these results in a BPS study to obtain the birthrates for
SNe Ia and the evolution of birthrates with time of SNe Ia for both a
constant star formation rate and a single star burst.  The birthrates
are somewhat lower than (but comparable to) those inferred
observationally.
 \end{abstract}


\section{Introduction}
Type Ia Supernovae (SNe Ia) appear to be good cosmological distance
indicators and have been applied successfully in determining
cosmological parameters (e.g. $\Omega$ and $\Lambda$; Riess et al.\
1998; Perlmutter et al.\ 1999).  They are believed to be thermonuclear
explosions of mass-accreting white dwarfs (WDs), although the exact
nature of their progenitors has remained unclear.  In the most widely
accepted model, the Chandrasekhar mass model, a carbon oxygen (CO) WD
accretes mass until it reaches a mass $\sim 1.378M_\odot$
\cite{nom84}, close to the Chandrasekhar mass, and explodes as a SN
Ia. One of the more popular scenarios in which this occurs is that a
CO WD accretes mass by mass transfer from a binary companion, which
may be a main sequence (MS) star or a slightly evolved subgiant.  This
channel is referred to as the WD+MS channel (van den Heuvel et al.\
1992; Rappaport, Di\,Stefano \& Smith 1994; Li \& van den Heuvel
1997).  Here the binary system originally consists of two MS stars,
where the more massive star evolves to become a CO WD by binary
interactions. Subsequentely, when the companion has evolved
sufficiently, it starts to fill its Roche lobe and to transfer
hydrogen-rich material onto the WD. The hydrogen accreted is burned
into helium, and then the helium is converted to carbon and
oxygen. The CO WD increases its mass till the mass reaches $\sim
1.378M_\odot$ when it explodes in a thermonuclear supernova.  Whether
the WD can grow in mass depends crucially on the mass-transfer rate
and the evolution of the mass-transfer rate with time.  If it is too
high, the system may enter into a common-envelope (CE) phase
(Paczy\'nski 1976), if it is too low, burning is unstable and leads to
nova explosions where all the accreted matter is ejected.

Hachisu et al.\ (1999, HKNU99), Hachisu, Kato \& Nomoto (1999, HKN99)
and Nomoto et al.\ \shortcite{nom99} have studied the WD+MS channel
for various metallicities.  However, their approach was based on a
simple analytical method for treating the binary interactions. It is
well established (e.g.\ Langer et al.\ 2000) that such analytical
prescriptions cannot describe certain mass-transfer phases, in
particular those occurring on a thermal timescale, appropriately.  Li
\& van den Heuvel \shortcite{li97} studied this channel (for a
metallicity $Z=0.02$) with detailed binary evolution calculations, but
only for two WD masses, $1.0\Ms$ and $1.2\Ms$, while Langer et al.\
\shortcite{lan00} investigated the channel (for metallicities $Z=0.02$
and 0.001) considering Case A evolution only (where mass transfer
occurs during core hydrogen burning).

The purpose of this paper is to study the binary channel more
comprehensively and to determine the detailed parameter range in which
this channel produces SNe Ia, which can be used for population
synthesis studies.  Employing the Eggleton stellar evolution code with
the latest input physics, we construct a fine grid of binary models for
a population I metallicity $Z=0.02$ in sections 2 and 3, and then
implement the results in a binary population synthesis (BPS) study in
section 4.  In section 5 we discuss the results and compare them to
previous studies and in section 6 we summarize the main results.

\section{Binary evolution calculation}

In the WD+MS channel, the lobe-filling star is a MS star or a star not
evolved far away from the MS.  The star transfers some of its mass
onto the WD, which grows in mass as a consequence. To determine
whether the WD reaches the critical mass for a SN Ia,
$1.378\Ms$, one needs to perform detailed binary evolution
calculations.  Here we use Eggleton's \shortcite{egg71,egg72,egg73}
stellar evolution code, modified to incorporate the WD accretion
prescription of HKNU99 into the binary calculations.

The Eggleton stellar evolution code has been updated with the latest
input physics over the last 3 decades, as described by Han et al.\
\shortcite{han94} and Pols et al.\ \shortcite{pol95,pol98}. Roche lobe
overflow (RLOF) is also treated within the code \cite{han00}.  In our
calculation, we use a typical Population I (Pop I) composition with
hydrogen abundance $X=0.70$, helium abundance $Y=0.28$ and metallicity
$Z=0.02$.  We set $\alpha =l/H_{\rm p}$, the ratio of typical mixing
length to the local pressure scaleheight, to 2, and set the convective
overshooting parameter $\delta_{\rm ov}$ to 0.12 \cite{sch97,pol97},
which roughly corresponds to an overshooting length of $\sim 0.25$
pressure scaleheights ($H_{\rm p}$).

We adopt the prescription of HKNU99 for the growth of the mass of a CO
WD by accretion of hydrogen-rich material from its companion.  If the
mass-transfer rate is above a critical rate, $\dot M_{\rm cr}$,
hydrogen burns steadily on the surface of the WD and, the hydrogen-rich
matter is converted into helium at a rate $\dot M_{\rm cr}$, while
the unprocessed matter is lost from the system, presumably in the
form of an optically thick wind.  The mass-loss rate of the wind is
therefore $\dot M_{\rm wind}=|\dot M_2| - \dot M_{\rm cr}$, where
$\dot M_2$ is the mass-transfer rate. The critical mass-transfer rate
is given by
\begin{equation}
\dot M_{\rm cr}=5.3\times 10^{-7}\, {(1.7-X)\over X} (M_{\rm WD}-0.4),
\end{equation}
where $X$ is the hydrogen mass fraction and $M_{\rm WD}$ the mass of the WD
(masses are in solar units and mass-accretion/transfer rates in
$\Ms\,\mbox{yr}^{-1}$).
If the mass-transfer rate is less than $\dot M_{\rm cr}$ but higher than
$\dot M_{\rm st}={1\over 2} \dot M_{\rm cr}$, it is assumed that there
is not mass loss and that 
hydrogen-shell burning is steady. If the mass-transfer rate is below
${1\over 2} \dot M_{\rm cr}$ but higher than ${1\over 8} \dot M_{\rm cr}$, 
hydrogen-shell burning is unstable, triggering
very weak shell flashes, where it is assumed that the processed
mass can be retained.
If the mass-transfer rate is lower than 
$\dot M_{\rm low}={1\over 8} \dot M_{\rm cr}$, hydrogen shell flashes are
so strong that no mass can be accumulated by the WD. Therefore
the growth rate of the mass of the helium layer on top of the CO WD is
\begin{equation}
\dot M_{\rm He}=\eta_{\rm H} |\dot M_2|,
\end{equation}
where 
\begin{equation}
\eta_{\rm H}=
\left\{ \begin{array}{l@{\quad,\quad}l}
\dot M_{\rm cr}\over |\dot M_2| &  |\dot{M_2}|>\dot{M}_{\rm cr},\strut\\
1 &  \dot{M}_{\rm cr}\ge |\dot{M_2}|\ge \dot{M}_{\rm low},\strut\\
0 &  |\dot{M_2}|< \dot{M}_{\rm low}.\strut\\
\end{array} \right.
\end{equation}
When the mass of the helium layer reaches a certain value, helium
ignites. If helium shell flashes occur, a part of the envelope mass is
blown off. The mass accumulation efficiency for helium shell flashes
according to HKNU99 is given by
\begin{equation}
\eta_{\rm He}=
\left\{ \begin{array}{l}
-0.175\, (\log \dot M_{\rm He}+5.35)^2+1.05, \\
\begin{array}{l@{\quad\quad\quad\quad\quad\quad}l}
& -7.3 < \log  \dot M_{\rm He} <-5.9, \\
1\quad , &  -5.9 \le \log  \dot M_{\rm He} \le -5.\\
\end{array} \end{array}\right.
\end{equation}
The growth rate of the CO WD is then given by
$\dot{M}_{\rm CO}$, i.e.
\begin{equation}
\dot{M}_{\rm CO}=\eta_{\rm He} \dot{M_{\rm He}}=
\eta_{\rm He} \eta_{\rm H}|\dot{M_2}|
\end{equation}
We assume that the mass lost from the system carries away the same specific
orbital angular momentum as the WD.

These prescriptions have been incorporated into our stellar evolution
code and allow us to follow the evolution of both the donor and the
accreting CO WD. Altogether, we have calculated the evolution of 2298
CO WD binary systems, thus obtaining a large, dense model grid. The
initial masses, $M_2^{\rm i}$, of donor stars range from 1.5 to
$4.0\Ms$, the initial masses, $M_{\rm WD}^{\rm i}$, of the CO
WDs from 0.67 to $1.2\Ms$, the initial orbital periods, $P^{\rm
i}$, of the binaries from the minimum period, at which a zero-age
main-sequence (ZAMS) star would fill its Roche lobe, to $\sim 30$ d.

\section{Binary Evolution Results}

\begin{figure*}
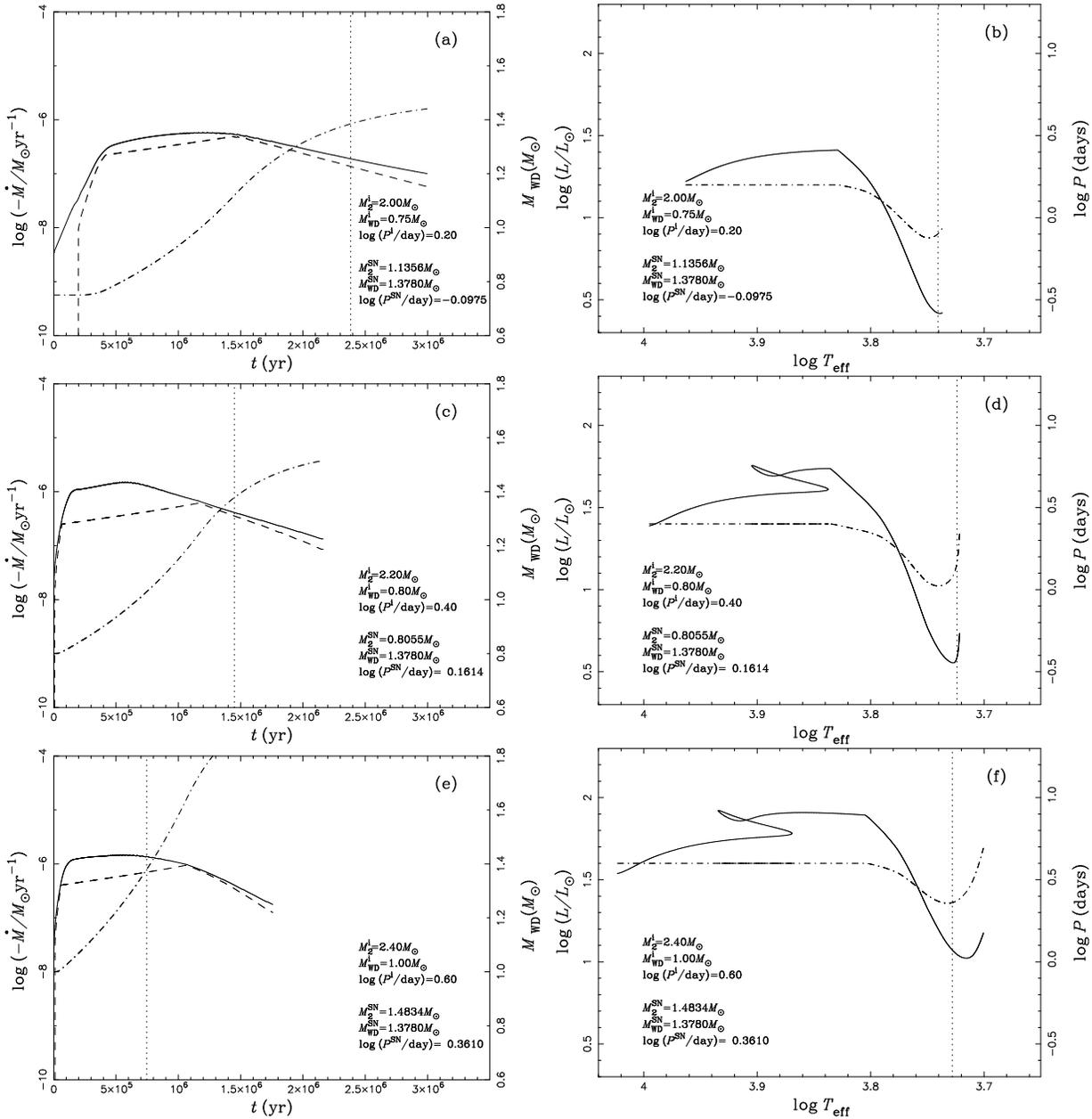

\centerline{\epsfig{file=mdot1.ps,angle=270,width=8cm}\ \
\epsfig{file=hrd1.ps,angle=270,width=8cm}}
\centerline{\epsfig{file=mdot2.ps,angle=270,width=8cm}\ \
\epsfig{file=hrd2.ps,angle=270,width=8cm}}
\centerline{\epsfig{file=mdot3.ps,angle=270,width=8cm}\ \
\epsfig{file=hrd3.ps,angle=270,width=8cm}}
\caption{Examples of three representative binary evolution
calculations. The solid, dashed, dash-dotted curves show the
mass-transfer rate, $\dot M_2$, the mass-growth rate of the CO WD,
$\dot M_{\rm CO}$, the mass of the CO WD, $M_{\rm WD}$, respectively,
in panels (a), (c) and (e). The evolutionary tracks of the donor stars
are shown as solid curves and the evolution of orbital period is shown
as dash-dotted curves in panels (b), (d) and (f). Dotted vertical
lines indicate the position where the WD is expected to explode in a
SN Ia in all panels. The initial binary parameters and the parameters
at the time of the SN Ia explosion are also given in each panel.  }
\label{mdot}
\end{figure*}
\begin{figure*}
\epsfig{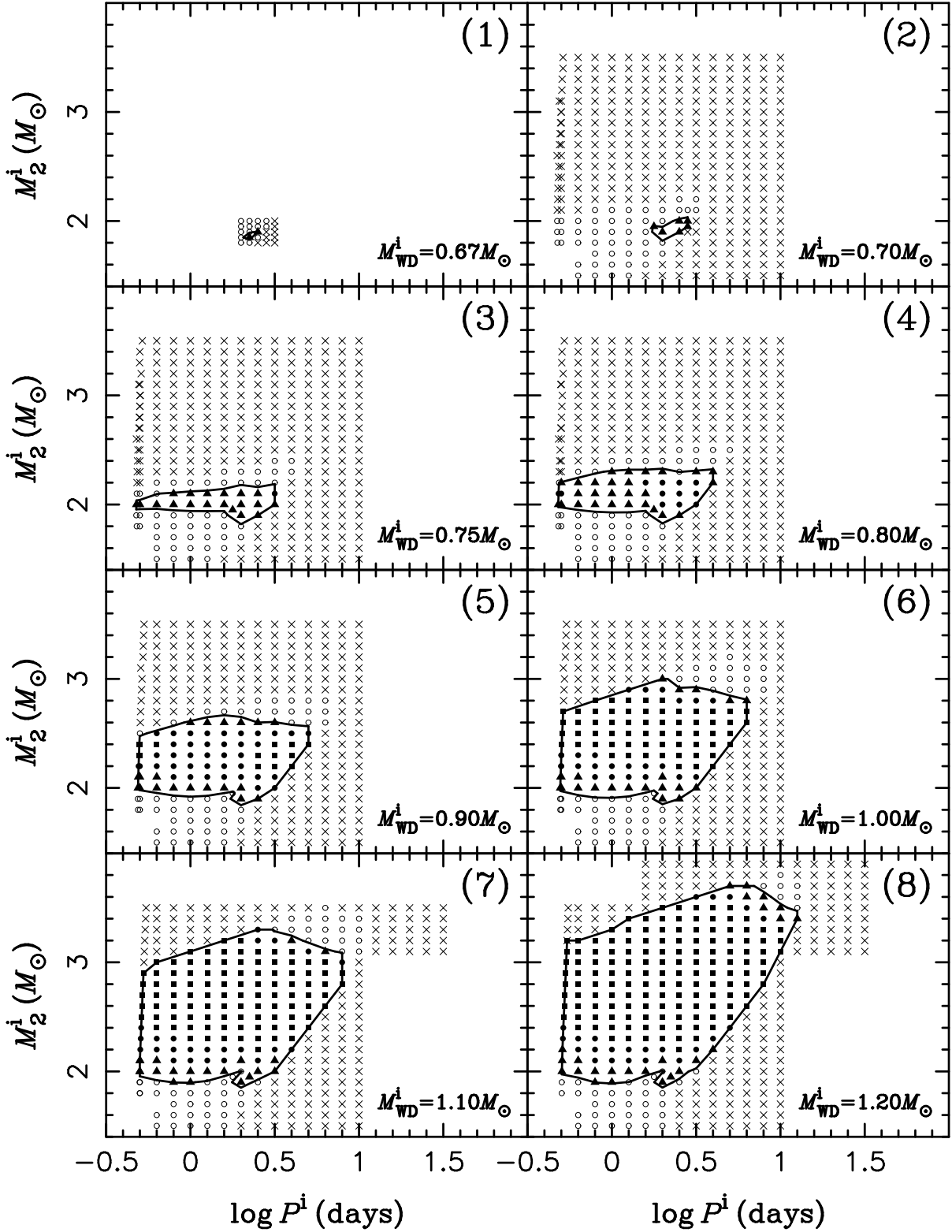}
\caption{Final outcomes of the binary evolution in the initial
orbital-period -- secondary mass ($\log P^{\rm i}$, $M^{\rm i}_2$)
plane of the CO+MS binary, where $P^{\rm i}$ is the initial orbital
period and $M^{\rm i}_2$ the initial mass of the donor star (for
different initial WD masses as indicated in each panel). Filled squares
indicate SN Ia explosions during an optically thick wind phase ($|\dot
M_2| \ge \dot M_{\rm cr}$), filled circles SN Ia explosions after the
wind phase, where hydrogen shell burning is stable ($\dot M_{\rm cr} >
|\dot M_2| \ge \dot M_{\rm cr}$), filled triangles Ia explosions after
the wind phase where hydrogen shell burning is mildly unstable ($\dot
M_{\rm st} > |\dot M_2| \ge \dot M_{\rm low}$). Open circles indicate
systems that experience novae, preventing the WD from reaching
$1.378\Ms$, while crosses show systems that are unstable to
dynamical mass transfer.  }
\label{grid}
\end{figure*}
\begin{figure}
\epsfig{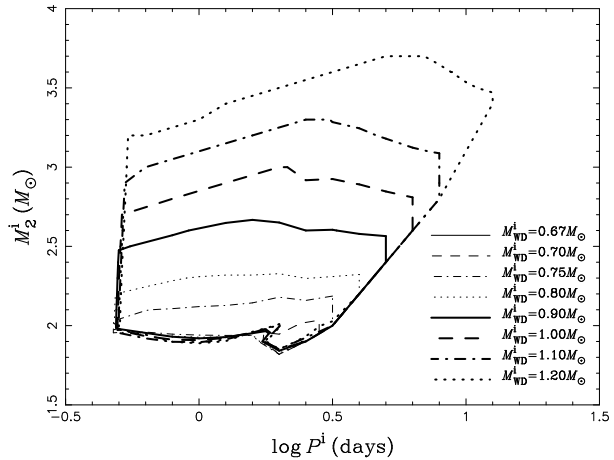}
\caption{
Regions in the initial orbital period -- secondary mass plane
($\log P^{\rm i}$, $M^{\rm i}_2$) for WD binaries that produce
SNe Ia for initial WD masses of
0.67, 0.70, 0.75, 0.80, 0.90, 1.0, 1.1 and $1.2\Ms$. The region
almost vanishes for $M_{\rm WD}^{\rm i}=0.67\Ms$.
}
\label{contour}
\end{figure}

In Figure~\ref{mdot} we present three representative examples of our
binary evolution calculations.  It shows the mass-transfer rate $\dot
M_2$, the growth rate of the CO WD, $\dot M_{\rm CO}$, the mass of the
CO WD, $M_{\rm WD}$, the evolutionary track of the donor star in the
Hertzsprung-Russel diagram (HRD) and the evolution of the orbital
period. Panels (a) and (b) represent the evolution of a binary system with
an initial mass of the donor star of $M_2^{\rm i}=2.00\Ms$, an
initial mass of the CO WD of $M_{\rm WD}^{\rm i}=0.75\Ms$ and an
initial orbital period of $\log (P^{\rm i}/{\rm day})=0.20$.  The
donor star fills its Roche lobe on the main sequence (MS) which
results in Case A RLOF. The mass-transfer rate exceeds $\dot M_{\rm
cr}$ soon after the onset of RLOF, leading to a wind phase, where part
of the transferred mass is blown off in an optically thick wind, while
the rest is accumulated by the WD.  When the mass-transfer rate drops
below $\dot M_{\rm cr}$ but it still higher than $\dot M_{\rm st}$,
the optically thick wind stops and hydrogen shell burning is stable.
The mass-transfer rate decreases further to below $\dot M_{\rm st}$
but remains above $\dot M_{\rm low}$, where hydrogen shell burning is
unstable, triggering very weak shell flashes, and the WD continues to
grow in mass.  When the mass reaches $M^{\rm SN}_{\rm
WD}=1.378\Ms$, the WD is assumed to explode as a SN Ia. At this
point, the mass of the donor is $M^{\rm SN}_2=1.1356\Ms$ and the
orbital period $\log (P^{\rm SN}/{\rm day})=-0.0975$.

Panels (c) and (d) of Figure~\ref{mdot} show another example for an
initial system with $M_2^{\rm i}=2.20\Ms$, $M_{\rm WD}^{\rm
i}=0.80\Ms$ and $\log (P^{\rm i}/{\rm day})=0.40$.  The binary
evolves in a similar way as in the previous example and the binary
parameters when the white dwarf reaches $M_{\rm WD}=1.378\Ms$ are
$M^{\rm SN}_2=0.8055\Ms$ and $\log (P^{\rm SN}/{\rm day})=0.1614$.
The main difference between this example and the previous one is that
RLOF begins in the Hertzsprung gap (i.e.\ after the secondary
has left the main sequence; so-called early Case B evolution) and
that, at the time of the explosion, 
the system is still in the stable hydrogen-burning phase after the
optically thick-wind phase.

Finally, the third example in panels (d) and (e) of Figure~\ref{mdot}
represents the case where mass transfer starts in the Hertzsprung gap
and where the binary remains in the optically-thick phase even at the
supernova stage. In this case, the initial binary parameters are
$M_2^{\rm i}=2.40\Ms$, $M_{\rm WD}^{\rm i}=1.00\Ms$ and $\log
(P^{\rm i}/{\rm day})=0.60$.  When $M_{\rm WD}=1.378\Ms$, they are
$M^{\rm SN}_2=1.4834\Ms$ and $\log (P^{\rm SN}/{\rm day})=0.3610$.

Figure~\ref{grid} shows the final outcome of the 2298 binary evolution
calculations in the initial orbital period -- secondary mass ($\log
P^{\rm i}$, $M^{\rm i}_2$) plane.  Filled symbols show that the
evolution leads to a SN Ia, where the shape of the symbols indicate
whether the WD explodes in the optically thick-wind phase (filled
squares; ($|\dot M_2| \ge \dot M_{\rm cr}$), after the wind phase in
the stable hydrogen-burning phase (filled circles: ($\dot M_{\rm cr} >
|\dot M_2| \ge \dot M_{\rm cr}$) or in the unstable hydrogen
shell-burning phase (filled triangles; ($\dot M_{\rm st} > |\dot M_2|
\ge \dot M_{\rm low}$). Systems which experience nova explosions and
never reach the Chandrasekhar limit and systems that may experience a
CE phase are also indicated in the figure.

In Figure~\ref{grid} and Figure~\ref{contour} we also present the
contours for the initial parameters for which a SN Ia results.  The
left boundaries of these contours in panels (3) to (8) are set by the
condition that RLOF starts when the secondary is still unevolved
(i.e.\ is on the ZAMS), while systems beyond the right boundary are
systems that experience dynamically unstable mass transfer when mass
transfer starts at the base of the red-giant branch (RGB). The upper
boundaries are also mainly set by the condition of dynamical
stability, since systems above the boundary have too large a mass
ratio for stable mass transfer.  The lower boundaries are caused by
the constraints that the mass-transfer rate has to be high enough for
the WD to be able to grow {\em} and that the donor mass is
sufficiently massive so that enough mass can be transferred to the WD to
reach the Chandrasekhar limit.  SNe Ia are either produced through
case A binary evolution (the left parts of the regions), in which
hydrogen is burning in the centre of the donor star at the onset of
RLOF, or early case B binary evolution (the right parts of the
regions), in which hydrogen is exhausted in the centre and the donor
is in the Hertzsprung gap at the onset of the RLOF phase.

In Figure~\ref{contour}, we overlay the contours for SN Ia production
in the ($\log P^{\rm i}$, $M^{\rm i}_2$) plane for initial WD masses of
0.67, 0.70, 0.75, 0.80, 0.90, 1.0, 1.1 and $1.2\Ms$. Note that the
enclosed region almost vanishes for $M_{\rm WD}^{\rm i}=0.67\Ms$,
which therefore sets the minimum WD mass for which this channel can
produce a supernova.  In Appendix~A, we present fitting formulae for
the boundaries of these contours (accurate to 3 per cent for most
boundaries), which can be used for population synthesis studies.  If
the initial parameters of a CO WD binary system is located in the SN
Ia production region, a SN Ia is assumed to be the outcome of the
binary evolution.

\section{Binary population synthesis}

In order to estimate the supernova frequencies in the WD+MS channel,
one first needs to determine the distribution and properties of
CO WD binaries. There are two channels to produce such 
systems.  In the first, the primary is in a relatively wide binary
and fills its Roche lobe as an asymptotic giant, where mass transfer 
is dynamically unstable. This leads to the formation of a common-envelope
(CE) phase (Pacz\'ynski 1976) and the spiral-in of the core of the
giant and the secondary inside this envelope (due to friction with the
CE).  If the orbital energy released in the orbital decay is able to
eject the envelope, this produces a rather close binary consisting of
the WD core of the primary (here, a CO WD) and the secondary.  As is
usually done in binary population synthesis (BPS) studies, we assume
that the CE is ejected if
\begin{equation}
\alpha_{\rm CE}\Delta E_{\rm orb}\ge |E_{\rm bind}|,
\end{equation}
where $\Delta E_{\rm orb}$ is the orbital energy released, $E_{\rm
bind}$ the binding energy of the envelope, and $\alpha_{\rm CE}$ the
common-envelope ejection efficiency, i.e.\ the fraction of the
released orbital energy used to overcome the binding energy.  We adopt
$E_{\rm bind}=E_{\rm gr}-\alpha_{\rm th} E_{\rm th}$, where $E_{\rm
gr}$ is the gravitational binding energy, $E_{\rm th}$ is the thermal
energy, and $\alpha_{\rm th}$ defines the fraction of the thermal
energy contributing to the CE ejection. 

A second channel to produce close CO WD binary systems involves
so-called case BB binary evolution \cite{del81}, 
where the primary fills its Roche
lobe when it has a helium core, leaving a helium star at the end of
the first RLOF phase. After it has exhausted the helium in the core,
the helium star, now containing a CO core, expands and fills its Roche
lobe again, transferring its remaining helium-rich envelope, producing
a CO WD binary in the process.

The CO WD binary system continues to evolve, and the secondary will at
some point also fill its Roche lobe; the WD will then start to accrete
mass from the secondary and convert the accreted matter into CO. We
assume that this ultimately produces a SN Ia if, at the beginning of
this RLOF phase, the orbital period, $P^{\rm i}_{\rm orb}$, and
secondary mass, $M_2^{\rm i}$, are in the appropriate regions in the
($\log P^{\rm i}$, $M_2^{\rm i}$) plane (see Figure~\ref{contour}) to
produce a SN Ia.

In order to investigate the birthrates of SNe Ia, we have performed a
series of detailed Monte Carlo simulations with the latest version of
the BPS code developed by Han et al.\ \shortcite{han03}.  In each
simulation, we follow the evolution of 100 million sample binaries
according to grids of stellar models of metallicity $Z=0.02$ and the
evolution channels described above.  We adopt the following input for
the simulations (see Han et al.\ 1995 for details).  (1) The
star-formation rate (SFR) is taken to be constant over the last 15\,Gyr
or, alternatively, as a delta function, i.e.\ a single star burst. In
the case of a constant SFR, we assume that a binary with its primary
more massive than $0.8\Ms$ is formed annually. For the case of a
single star burst, we assume a burst producing $10^{11}\Ms$ in
stars.  (2) The initial mass function (IMF) of Miller and Scalo
\shortcite{mil79} is adopted. (3) The mass-ratio distribution is taken
to be constant or, alternatively, it is assumed that the component
masses are uncorrelated. (4) We take the distribution of
separations to be constant in $\log a$ for wide binaries, where $a$ is
the orbital separation.  Our adopted distribution implies that $\sim
50$\,\% of stellar systems are binary systems with orbital periods
less than 100 yr.

\begin{figure}
\epsfig{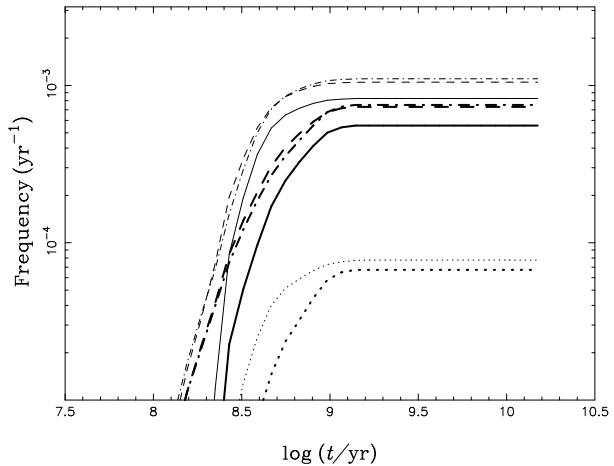}
\caption{
The evolution of birthrates of type Ia supernovae (SNe Ia) for
a constant Pop I star-formation rate ($3.5\Ms\,$yr$^{-1}$).
Solid, dashed, dash-dotted curves are for a constant initial
mass-ratio distribution and with
$\alpha_{\rm CE}=\alpha_{\rm th}=1.0$ (solid), $0.75$ (dashed) and
$0.5$ (dash-dotted), respectively. Dotted curves are with
$\alpha_{\rm CE}=\alpha_{\rm th}=1.0$ and 
for a mass-ratio distribution
with uncorrelated component masses.
Thick curves are based on the contours in this paper, while thin
curves are based on the results of Hachisu et al.\ (1999).
}
\label{rate-c}
\end{figure}

\begin{figure}
\epsfig{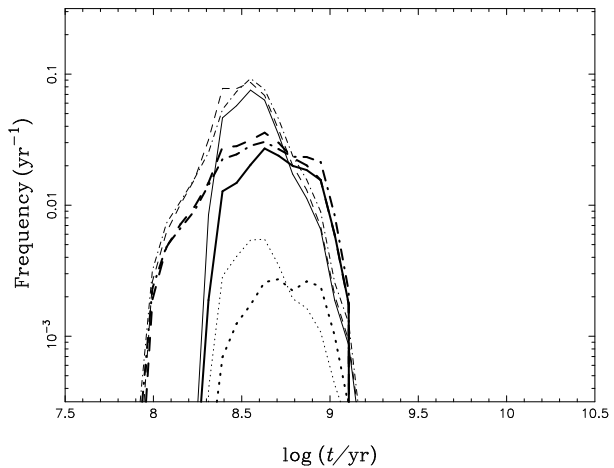}
\caption{
Similar to Figure~\ref{rate-c}, but for a single star burst of
$10^{11}\Ms$.
}
\label{rate-s}
\end{figure}

\begin{figure}
\epsfig{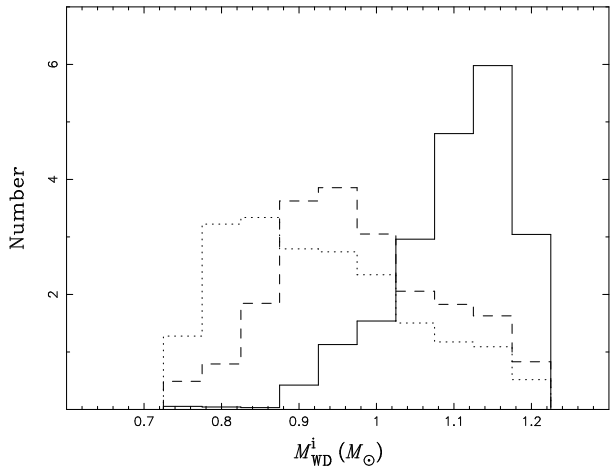}
\caption{ The distribution of the initial masses of the CO WDs for the
progenitors of SNe Ia according to the contours in this paper. The
solid, dashed and dotted histograms are for SNe Ia produced at 0.3,
0.5 and 0.7\,Gyr after a single star burst. The simulation uses a
metallicity $Z=0.02$, a constant initial mass-ratio distribution and
$\alpha_{\rm CE}=\alpha_{\rm th}=1.0$.  }
\label{mwdi-h}
\end{figure}

\begin{figure}
\epsfig{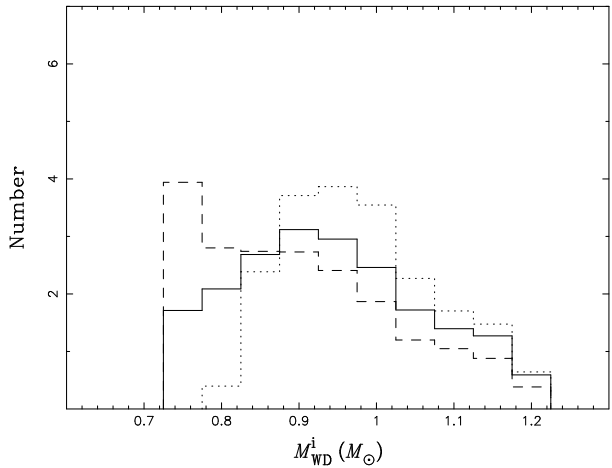}
\caption{
Similar to Figure~\ref{mwdi-h}, but based on the
contours of Hachisu et al. (1999). 
}
\label{mwdi-n}
\end{figure}

The results of the simulations are plotted in 
Figures~\ref{rate-c}-~\ref{mwdi-n}.

\section{Discussion}

\begin{figure}
\centerline{\epsfig{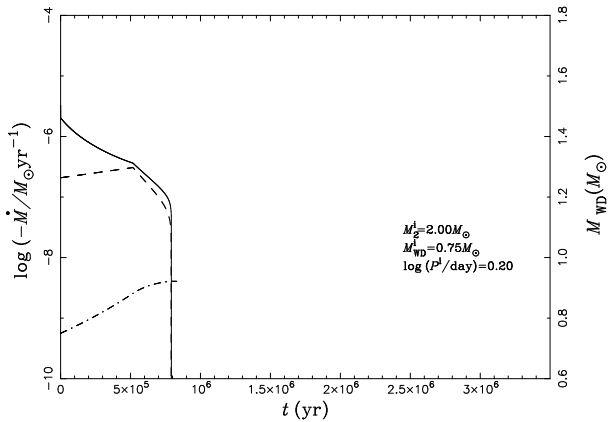}}
\caption{ The evolution of mass-transfer rate, $\dot{M}_2$,
mass-growth rate of the CO WD, $\dot{M}_{\rm CO}$, and the CO WD mass,
$M_{\rm WD}$, similar to panel (a) of Figure~\ref{mdot}, but using an
analytical approach similar to the one of Hachisu et al.\ (1999)
rather than full binary calculations.}
\label{mdot-nom}
\end{figure}

\begin{figure}
\epsfig{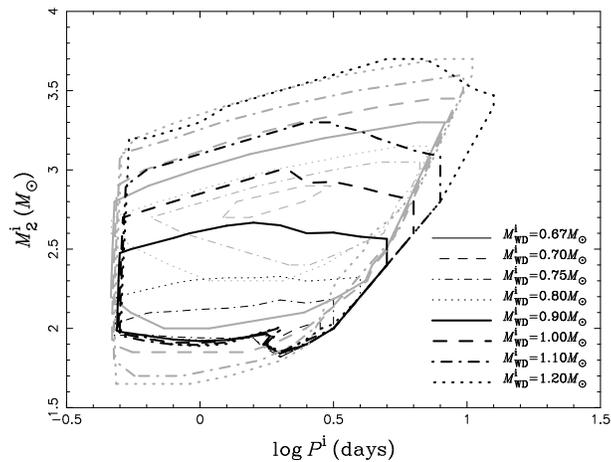}
\caption{ A comparison of the results of this paper with those
obtained by Hachisu et al.\ (1999, HKNU99) and Hachisu, Kato and
Nomoto (1999, HKN99). Dark contours show the parameter regions in the
($\log P,\log M_{2}^{\rm i}$) plane for different white-dwarf masses
that lead to a SN Ia. The light grey contours are taken from HKN99 and
HKNU99.  }
\label{nomoto}
\end{figure}


\subsection{Comparisons with previous studies}

Langer et al.\ \shortcite{lan00} considered the CO+MS channel
considering case A evolution for metallicities $Z=0.02$ and $Z=0.001$,
respectively.  They used a rather simple prescription for the WD mass
accretion.  We checked that we obtain very similar results,
e.g. similar mass-transfer rates, similar final WD masses, if we adopt
similar prescriptions.  They also pointed out the possibility that
binary systems with rather small initial white dwarf masses ($\sim
0.7\Ms$) may produce SNe Ia. This agrees with our results
that show that a binary system with $M_{\rm WD}^{\rm i}=0.67\Ms$
can produce an SN Ia.  Li \& van den Heuvel \shortcite{li97} studied
the evolution of WD binaries in search for progenitors of SNe Ia by
using assumptions similar to those of Hachisu, Kato \& Nomoto (1996;
1999) and Hachisu et al.\ (1999), but only for two WD masses, 1.0 and
$1.2\Ms$. Our contours for the regions of WD binaries producing
SNe Ia are consistent with theirs if we take into account that their
model grid is much smaller and that their stellar evolution model
parameters (e.g. overshooting parameter) may be different from ours.

Hachisu et al.\ (1999, HKNU99), Hachisu, Kato and Nomoto (1999, HKN99)
and Nomoto et al.\ \shortcite{nom99} studied the WD+MS channel for
progenitors of SNe Ia with different metallicities. They used the analytic
fitting formulae of Tout et al.\ \shortcite{tou97} to calculate the
radius and the luminosity of a main-sequence star. As the mass
transfer proceeds on a thermal timescale for $M_2^{\rm i}/M_{\rm
WD}^{\rm i}>0.79$, they approximated the mass-transfer rate as
\begin{equation}
|\dot M_2|={M_2\over \tau_{\rm KH}}{\rm Max}
\Bigl( {\zeta_{\rm RL}-\zeta_{\rm MS}\over \zeta_{\rm MS}}, 0\Bigr)
\end{equation}
where $\tau_{\rm KH}$ is the Kelvin-Helmholtz timescale, and
$\zeta_{\rm RL}$ and $\zeta_{\rm MS}$ are the mass-radius exponents of
the inner critical Roche lobe and the main-sequence star,
respectively.  In order to be able to make a comparison, we adopted a
similar approach to calculate the evolution of a binary system with
initial parameters $M_2^{\rm i}=2.00\Ms$, $M_{\rm WD}^{\rm
i}=0.75\Ms$ and $\log (P^{\rm i}/{\rm day})=0.20$, the results
of which are shown in Figure~\ref{mdot-nom}. Comparison of
Figure~\ref{mdot-nom} and panel (a) of Figure~\ref{mdot} shows that
the estimate of the mass-transfer rate is significantly different in
this model compared to the results of our detailed binary calculations
(also see Langer et al.\ 2000 for a detailed analysis).  The
estimated mass-transfer rate is significantly larger, and the
mass-transfer phase is correspondingly much shorter. The contours of
HKNU99 are plotted as light grey curves in Figure~\ref{nomoto}.  Our
contours are similar to theirs for massive WDs, but very different for
low-mass WDs. For example, the contour for $M^{\rm i}_{\rm
WD}=0.75\Ms$ in HKNU99 corresponds to significantly more massive
donors ($\sim 2.7\Ms$) than our contour ($\sim 2.0\Ms$).  The
enclosed region of the contour of HKNU99 is also larger. We consider
our contours more physical as they are based on full binary evolution
calculations rather than a simple prescription for estimating the
mass-transfer rate and use the latest stellar evolution physics.

\subsection{Birthrates of SNe Ia}

Figure~\ref{rate-c} shows Galactic birthrates of SNe Ia for the WD+MS
channel.  The simulations for a constant initial mass-ratio
distribution give a Galactic birthrate of $0.6-1.1\times 10^{-3}\ {\rm
yr}^{-1}$, lower but not dramatically lower than those inferred
observationally (i.e.\ within a factor of a few: $3 - 4\times 10^{-3}\
{\rm yr}^{-1}$: van den Bergh \& Tammann, 1991; Capellaro \& Turatto
1997).  The simulation for the initial mass-ratio distribution with
uncorrelated component masses gives a birthrate that is lower by an
order of magnitude, as most of the donors in the WD+MS channel are not
very massive which has the consequence that the white dwarfs cannot
accrete enough mass to reach the Chandrasekhar limit.  The birthrates
based on our new contours (Fig.~\ref{contour}) are
$0.6-0.8\times 10^{-3}\ {\rm yr}^{-1}$, somewhat lower than the
birthrates one would obtain whith the contours of HKNU99 for a
constant mass-ratio distribution ($0.8-1.1\times 10^{-3}\ {\rm
yr}^{-1}$). This follows directly from the fact that the parameter
regions where low-mass WDs can produce a SN Ia is smaller in our study
than in HKNU99 (see Fig.~\ref{nomoto}).

Yungelson \& Livio \shortcite{yun98} gave a much lower SN Ia birthrate
for the WD+MS channel, which is just 10\% of the inferred SN Ia rate.
HKN99 speculated that an important evolutionary path (case BB
evolution) was not included in the model of Yungelson \& Livio
\shortcite{yun98}. However, Yungelson (private communication) has since
pointed out that case BB evolution was indeed included though not
mentioned especially. Since we have also included this channel, we
suspect that the differences between our result and theirs may be due
to different treatments of CE evolution (see the discussion of Han,
Podsiadlowski, Eggleton 1995) and perhaps most significantly
different assumptions concerning the efficiency of hydrogen accumulation,
where their efficiency is lower than ours.

HKN99 proposed a wide symbiotic channel (i.e. WD+RG channel, in which
a low-mass red giant is the mass donor) to increase the number of
possible SN Ia progenitors and gave the corresponding regions in the
($\log P^{\rm i}$, $M_2^{\rm i}$) plane. However, at present it is not
clear from a BPS point of view how to populate these parameter regions
with WD systems. Our BPS model would predict that the Galactic SN Ia rate
from the WD+RG channel is negligible ($\sim 2\times 10^{-5}\ {\rm
yr}^{-1}$). This is consistent with the conclusions of Yungelson \&
Livio \shortcite{yun98}, although we think that this needs to be
investigated further.

Figure~\ref{rate-s} displays the evolution of birthrates of SNe Ia for
a single star burst of $10^{11}\Ms$. Most of the supernova
explosions occur between 0.1 and 1\,Gyr after the burst.  The figure
also shows that a high common-envelope ejection efficiency leads to a
systematically later explosion time, since a high CE ejection
efficiency leads to wider WD binaries and it takes a longer time for
the secondary to evolve to fill its Roche lobe.

\subsection{The distribution of the initial WD masses}

Figure~\ref{mwdi-h} shows the distributions of the initial masses of
the CO WDs that ultimately produce a SNe Ia according to our new
calculations. The distributions are shown for systems that produce SNe
Ia at 0.3, 0.5 and 0.7\,Gyr after a single star burst. For clarity, we
only show the distributions for the simulation with a constant initial
mass-ratio distribution and for $\alpha_{\rm CE}=\alpha_{\rm th}=1.0$,
as the other simulations gave similar results. The figure shows that
there is clear trend for the peak in the initial WD mass distribution
to move to lower masses with time, where the first SNe Ia
preferentially have massive WD progenitors.  Such a clear trend is not
seen when the contours of HKNU99 are used (Fig.~\ref{mwdi-n}). Here
the trend is not so clear, and the peak moves first to lower mass and
subsequently to higher mass as the age increases. The difference
between Figure~\ref{mwdi-h} and Figure~\ref{mwdi-n} can be entirely
understood from the different behaviour of the contours in the two
studies and the fact that a massive donor in the WD+MS channel will
evolve more quickly and hence produces a supernova at an earlier time.

The brightness of a SN Ia is mainly determined by the mass of
$^{56}{\rm Ni}$ synthesized ($M_{\rm Ni}$) during the explosion. As
pointed out by H\"oflich, Wheeler \& Thielemann (1997) and Nomoto et
al.\ \shortcite{nom99}, the amount of $^{56}$Ni synthesized will
generally depend on the C/O ratio in the core of the WD just before the
explosion. Nomoto et al.\ (1999) speculated that the supernova
brightness increases with the C/O ratio.  We tested the C/O relation
with mass in our stellar evolution code and found, in agreement with
earlier studies, that the C/O ratio decreases with increasing ZAMS
mass and hence white dwarf mass.  This would suggest that older
progenitors, which tend to have lower initial WD mass
(Fig.~\ref{rate-c}), should produce brighter SNe Ia in the WD+MS
channel. However, irrespective of what the correlation of the C/O
ratio with explosion energy may be, it is clear that the C/O ratio is
important in determining the explosion energy. Since the C/O ratio
depends on the age of the population and also the metallicity (see
H\"oflich et al.\ 1997, Langer et al.\ 2000 for further discussions),
this may introduce evolutionary effects that may need to be taken into
account in cosmological applications of SNe Ia.

Nomoto et al.\ \shortcite{nom99} have studied the distribution of the
initial WD masses both for the WD+MS channel and the WD+RG channel.
They found that the WD+RG channel tends to have higher WD masses on
average as a more massive WD is generally needed to allow stable RLOF
(rather than lead to dynamical mass transfer and a common-envelope
phase). As a consequence, SNe Ia from the WD+RG channel would tend to
be dimmer (assuming the above relation between supernova brightness
and C/O ratio/WD mass).  At an age of 10\,Gyr, the WD+RG channel is
the major single-degenerate channel to produce SNe Ia. This could
explain why SNe Ia are much dimmer in elliptical galaxies. In our BPS
simulations, the WD+RG channel was not important (see the earlier
discussion), and the WD+MS channel alone cannot explain the SNe Ia
in elliptical galaxies. We will investigate this issue further,
although at face value it might suggest that the theoretically less
favoured double degenerate (DD) channel for SNe Ia \cite{ibe84,web87},
in which two CO WDs with a total mass larger than the Chandrasekhar
mass limit coalesce, may be required \cite{yun98,han98,yun00,tut02}.

Finally, Langer et al. \shortcite{lan00} and Nomoto et al.\
\shortcite{nom99} pointed out that a more massive WD is needed to
produce a SN Ia at low metallicity. As a low metallicity is in some
sense similar to a young age from a stellar evolution point of view,
this implies that the SNe Ia in low-metallicity populations should be
dimmer.

\section{Summary and Conclusion}
Adopting the prescription of HKNU99 for the mass accretion of CO WDs,
we have carried out detailed binary evolution calculations for the
progenitors of SNe Ia in the semi-degenerate channel (the WD+MS
channel) and obtained the initial parameters in the ($\log P^{\rm i}$,
$M_2^{\rm i}$) plane that lead to SNe Ia.  We provided fitting
formulae that give the contours for the initial binary parameters that
can be used in BPS studies.  We find that a CO WD with its mass as low
as $0.67\Ms$ can accrete mass and reach the Chandrasekhar mass
limit.  By incorporating the contours into our BPS code, we obtained
the birthrates of SNe Ia and their evolution with time.  The Galactic
birthrate is lower than but still comparable within a factor of
a few to that inferred observationally.

\section*{Acknowledgements}
We thank Prof.\ Yungelson for useful discussions.
This work was in part supported by a Royal Society UK-China Joint Project
Grant (Ph.P and Z.H.), the Chinese National Science Foundation under 
Grant No.\ 19925312, 10073009 and NKBRSF No. 19990754 (Z.H.) and 
a European Research \& Training Network on Type Ia Supernovae
(HPRN-CT-20002-00303).

\onecolumn
\appendix
\section{Fitting formulae for the boundaries of SN Ia production regions}

The upper boundaries of the contours in Figure~\ref{contour} can be fitted as
\begin{eqnarray}
      M_2^{\rm i}={\rm Min}(A+B\log P^{\rm i} +C(\log P^{\rm i})^2
      +D(\log P^{\rm i})^3,\ 3.2M_{\rm WD}^{\rm i})
\label{upper}                
\end{eqnarray}
where
\begin{equation}
      \left\{ \begin{array}{l}
      A=-1.9+7.078M_{\rm WD}^{\rm i}-2.284(M_{\rm WD}^{\rm i})^2\\
      B=-1.482+2.504M_{\rm WD}^{\rm i}-0.5917(M_{\rm WD}^{\rm i})^2\\
      C=13.88-31.50M_{\rm WD}^{\rm i}+16.86(M_{\rm WD}^{\rm i})^2\\
      D= 1.205+1.243M_{\rm WD}^{\rm i}-2.414(M_{\rm WD}^{\rm i})^2\\
      \end{array}\right.
\end{equation}
In the fitting we ignored the vertical parts (at the right of the upper
boundaries) as they are due to the finite spacing of our model
grid.

The lower and the right boundaries can be fitted as
\begin{equation}
      M_2^{\rm i}=\left\{\begin{array}{ll}
          1.946-0.03261M_{\rm WD}^{\rm i}-0.03639\log P^{\rm i}
                  +0.4366(\log P^{\rm i})^2 &
                 {\qquad\qquad} {\rm if} \log P^{\rm i}\le 0.24\\
          1.145-0.007195M_{\rm WD}^{\rm i}+
                  0.02563/(\log P^{\rm i})^2+ 1.175\log P^{\rm i}
                  +0.7361(\log P^{\rm i})^2 &
      {\qquad\qquad} {\rm if} \log P^{\rm i}> 0.24\\
\end{array}\right.
\label{lower}
\end{equation}

The left boundaries are either defined by the onset of RLOF from the
secondary on the ZAMS for high-mass
WDs, or by the onset of RLOF at the beginning of the
Hertzsprung gap for low-mass WDs. We also carried out additional binary evolution
calculations and found that, when $M_{\rm WD}^{\rm i}$ is lower than
$0.73\Ms$, SNe Ia cannot be produced by case A evolution. We 
therefore obtain
\begin{equation}
      \log P^{\rm i}=\left\{ \begin{array}{ll}
       -0.4096+ 0.05301M_{\rm WD}^{\rm i}+ 0.02488M_2^{\rm i},
        & {\qquad\qquad} {\rm if} M_{\rm WD}^{\rm i}\ge 0.73\\
      0.24,& {\qquad\qquad} {\rm if} M_{\rm WD}^{\rm i}<0.73\\
      \end{array}\right.
\label{left}
\end{equation}

In the above fitting formulae, the masses are in \Ms\ and the
orbital period is in days. The formulae are valid for $0.6\Ms \le
M_{\rm WD}^{\rm i} \le 1.2\Ms$, $1.5 \Ms \le M_2^{\rm i} \le
4.0\Ms$ and $-0.5 \le \log (P^{\rm i}/{\rm day}) \le 1.5$.  If the
initial parameters of a CO WD binary system is located in the SN Ia
production region, i.e. the binary is below the upper boundary
(equation~\ref{upper}), above the lower and the right boundaries
(equation~\ref{lower}) and to the right of the left boundary
(equation~\ref{left}), a SN Ia is assumed to result from the binary
evolution.  The formulae are written into a FORTRAN code and the code
can be supplied on request by contacting Z.H.

\end{document}